\documentclass[11pt]{article}

\usepackage[margin=1in]{geometry}
\usepackage{amsmath,amssymb,amsthm,mathtools}
\usepackage{bm}
\usepackage{comment}
\usepackage{lipsum}
\usepackage[T1]{fontenc}
\usepackage[colorlinks=true,citecolor=blue,linkcolor=blue,urlcolor=blue]{hyperref}
\usepackage{caption}
\usepackage{subcaption}
\usepackage{xcolor}

\usepackage{enumitem}

\usepackage[authoryear]{natbib}

\usepackage{tabularx}

\renewenvironment{quote}{
	\list{}{
		\leftmargin0.5cm 
		\rightmargin\leftmargin
	}
	\item\relax
}
{\endlist}

\usepackage[ruled,vlined,linesnumbered]{algorithm2e}
\makeatletter
\newcommand{\algorithmfootnote}[2][\footnotesize]{%
	\let\old@algocf@finish\@algocf@finish%
	\def\@algocf@finish{\old@algocf@finish%
		\leavevmode\rlap{\begin{minipage}{\linewidth}
				#1#2
		\end{minipage}}%
	}%
}
\makeatother

\usepackage{graphicx}
\graphicspath{{images/}}

\newtheorem{theorem}{Theorem}
\newtheorem{definition}{Definition}
\newtheorem{assumption}{Assumption}
\newtheorem{proposition}{Proposition}
\newtheorem{example}{Example}

\newcommand{\cX}{\mathcal{X}}
\newcommand{\cZ}{\mathcal{Z}}
\newcommand{\cE}{\mathcal{E}}
\newcommand{\cB}{\mathcal{B}}
\newcommand{\cP}{\mathcal{P}}

\newcommand{\cN}{\mathcal{N}}
\newcommand{\cF}{\mathcal{F}}
\newcommand{\cY}{\mathcal{Y}}
\newcommand{\bfb}{\mathbf{b}}
\newcommand{\bx}{\mathbf{x}}
\newcommand{\bv}{\mathbf{v}}
\newcommand{\bbR}{\mathbb{R}}

\newcommand{\bz}{\mathbf{z}}
\newcommand{\by}{\mathbf{y}}
\newcommand{\iid}{\stackrel{\scriptscriptstyle{\text{i.i.d.}}}{\sim}}

\title{A User-Driven Framework for\\ Regulating and Auditing Social Media}

\author{Sarah H. Cen\\
	  \hspace{4pt}   \texttt{shcen@mit.edu}  \hspace{4pt}  \\
	MIT
	\and 
	Aleksander M\k{a}dry \\
	\texttt{madry@mit.edu} \\
	MIT
\and Devavrat Shah \\
\texttt{devavrat@mit.edu} \\
MIT}

\date{}

\begin{document}

\maketitle
\begin{abstract}

People form judgments and make decisions based on the information that they observe. 
A growing portion of that information is not only provided, but carefully curated by social media platforms. 
Although lawmakers largely agree that platforms should not operate without any oversight, 
there is little consensus on how to regulate social media. 
There is consensus, however, that creating a strict, global standard of ``acceptable'' content is untenable (e.g., in the US, it is incompatible with Section 230 of the Communications Decency Act and the First Amendment). 

In this work, we propose that algorithmic filtering should be regulated with respect to a \emph{flexible, user-driven baseline}. 
We provide a concrete framework for regulating and auditing a social media platform according to such a baseline. 
In particular, 
we introduce the notion of a \emph{baseline feed}: 
the content that a user would see without filtering (e.g., on Twitter, this could be the chronological timeline).  
We require that the feeds a platform filters contain ``similar'' informational content as their respective baseline feeds,
and we design a principled way to measure similarity.
This approach is motivated by related suggestions that regulations should increase user agency. 
We present an auditing procedure that checks whether a platform honors this requirement. 
Notably, the audit needs only black-box access to a platform's filtering algorithm, and it does not access or infer private user information.
We provide theoretical guarantees on the strength of the audit. 
We further show that requiring closeness between filtered and baseline feeds does \emph{not} impose a large performance cost, nor does it create echo chambers. 

\end{abstract}

\section{Introduction}\label{sec:intro}

Despite a growing desire to regulate social media, there has been little consensus on \emph{how} and \emph{to what extent} social media should be regulated.

This conversation dates back several years, as the public began to realize that social media platforms do not simply provide, but rather carefully curate (or \emph{filter}) the content that appears on each user’s feed. Compounded by the fact that social media enjoys a user base of nearly five billion, the ability to \emph{personalize} content places significant power in the hands of platforms. By shaping the lens through which users receive information, platforms can influence macro-decisions, such as how a nation votes, as well as micro-decisions, such as what we choose to eat, watch, buy, and more \citep{sidani2016association,garvin2019shopping,tucker2018social,chen2019understanding,broughton2013use,propub2017housing}. 

In response, many began to wonder whether platforms should be able to curate information on such massive scales without oversight. For the most part, policymakers share a desire to regulate, pointing to evidence that, left to their own devices, social media platforms do not always act in the public’s best interest (e.g., that Facebook amplified misinformation during the 2016 US elections \citep{allcott2019trends}—in some cases preventing voters from getting to the polls—and that Instagram knowingly promoted content that is toxic for teenage mental health \citep{wells2021facebook}). 

However, in the US, progress on social media regulation has largely stalled. Why? For one, Section 230 of the Communications Decency Act declares that platforms cannot be treated as “publishers,” drawing a clear line between platforms and traditional media sources (e.g., newspapers) \citep{brannon2021section}. For another, the US Supreme Court has historically struck down regulations that restrict free speech (even if these regulations do so indirectly), and many of the popular proposals (e.g., outlawing “harmful” content) will likely receive the same treatment \citep{keller2021amplification,1959smith}. In addition, policymakers are worried that regulation may stifle innovation and interfere with personalization (the very feature that users love).

In response, several works have signaled that there is an alternative approach to social media regulation that does not clash with the legislative considerations mentioned above \citep{vese2022governing,balkin2021regulate,keller2021amplification}. Broadly speaking, these works argue against regulating with respect to a strict standard of “acceptable” content (e.g., requiring that platforms remove misinformation, or requiring that half of every user’s election-related content is left-leaning and half is right-leaning). They maintain that such global baselines are often brittle, asserting the need for a \emph{flexible} baseline. 

In this work, we propose a way to formulate such a flexible baseline. We refer to this approach as \textbf{regulating with respect to a user-driven baseline} and propose a concrete framework for {regulating with respect to this baseline}. This framework gives rise to a procedure for auditing a platform’s content filtering algorithm. The audit tests whether the platform complies with the user-driven baseline and, notably, does not require disclosing the specifics of (i.e., white-boxing) the algorithm. Below, we describe our contributions in more detail.

\begin{figure*}[t]
	\includegraphics[width=\textwidth]{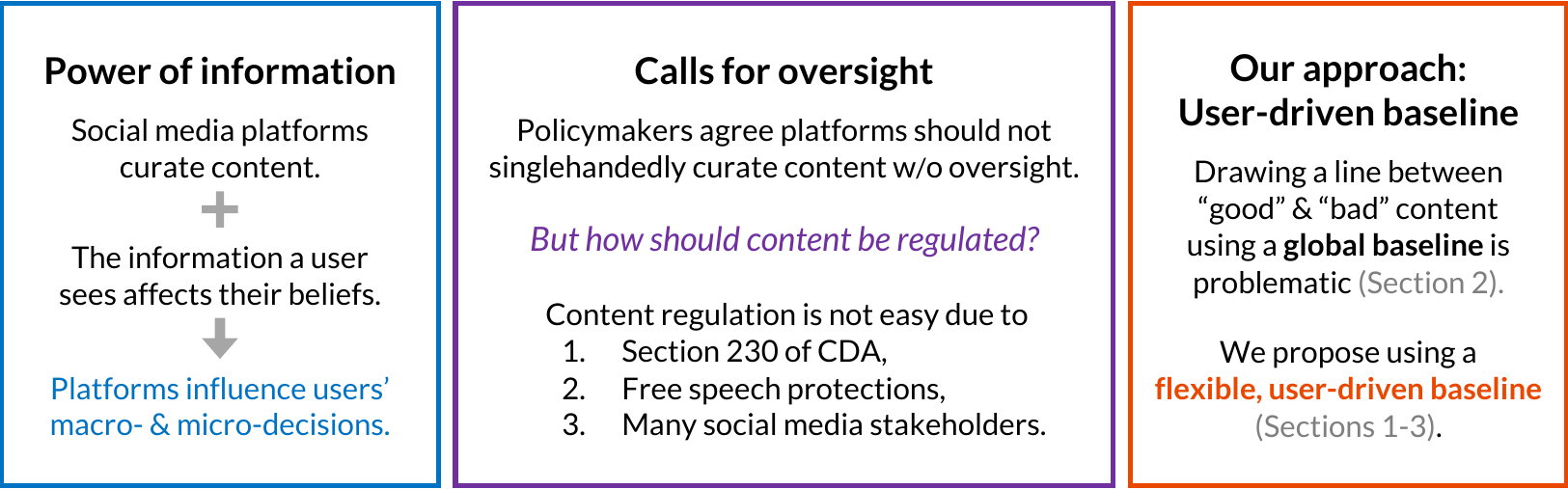}
	\centering
	\caption{By curating the content that appears on each user's feed, social media platforms wield an enormous amount of influence. Although policymarkers largely agree that platforms should be subject to some oversight, it is unclear how and to what extent social media should be regulated. In recent years, many have found that using a strict, \emph{global} baseline to draw a line between ``acceptable'' and ``unacceptable'' content is problematic. In this work, we propose using a flexible, user-driven baseline, as described in Sections \ref{sec:intro}-\ref{sec:problem_statement} and Figure \ref{fig:decision_robustness}.}
	\label{fig:motivation}
\end{figure*}

\paragraph{Our contributions.} 
We focus our attention on the regulation of \emph{algorithmic filtering}: the process by which platforms curate the content on each user’s feed. To regulate the effect of algorithmic filtering, we first quantify its effect on users using the following intuition. 

In many cases, there is a notion of a \emph{baseline} (i.e., natural) feed. On Twitter, for example, the baseline feed could be the chronological feed—a stream of Tweets from accounts that a user has chosen to follow, given in reverse chronological order \citep{huszar2022algorithmic,bartley2021auditing,bandy2021more}. On other platforms, the baseline feed could correspond to a mix of content from the users’ friends and pages that she has followed. 

The key principle underlying our regulatory framework is that the content a platform filters for a user must be “close” a baseline feed for the same user. This requirement reflects the reasoning that algorithmic filtering should not be regulated with respect to a strict, global standard chosen by lawmakers or platforms, as such an approach would reduce the already diminished amount of power that users have. Rather, \emph{algorithmic filtering should be regulated with respect to a flexible, user-driven standard, i.e., the baseline feed}. As further discussed in Section \ref{sec:background}, this approach can be justified in several ways, including as a mechanism for increasing user agency. 

Although such an approach “would not prevent individuals actively choosing lawful but harmful or polarizing content online, anymore than current law prevents the same choice in media consumption,” it would ensure that platforms—like Facebook and Twitter—do not singlehandedly choose what millions of users see \citep{keller2021amplification}. Additionally, as \citet{keller2021amplification} puts it, this approach does not violate Section 230 or free speech protections because “internet users who lost reach and audience under such a law would have little basis to object, since that loss would stem from other users’ choice not to listen to them”  (cf. Section \ref{sec:background} for more detail). 
\\

\noindent 
Requiring that the filtered feed be close to a baseline feed gives rise to two important questions.
 
First, \textbf{what is an appropriate way to measure the distance between a filtered feed and a  baseline feed}? For example, what should it mean for a user's filtered Twitter feed to be “close” to their chronological feed? 
One naive approach would be to threshold the $\ell_p$-norm of the difference between the filtered and baseline feeds, but it is unclear that thresholding the $\ell_p$ norm is meaningful, nor is it clear how one should choose the threshold in a principled manner. 

Second, \textbf{is requiring that filtered feeds be close to their respective baseline feeds too restrictive}? Specifically, does it prevent platforms from being profitable or hurt their ability to personalize (a service that users covet)? Or does it cause echo chambers? One might expect that if the “closeness” requirement is too stringent, both users and platforms are prevented from reaping the benefits of algorithmic filtering. 

Our framework addresses both these questions. To address the first, we use the notion of \emph{decision robustness} proposed by \citet{cen2021regulating} as a meaningful measure of the distance between two feeds. We then provide an auditing procedure that checks whether a platforms' filtering algorithm is decision-robust. The audit requires only black-box access to the platform's filtering algorithm and does not need access to the users' personal data.  In answer to the second question, we examine how the audit affects the platform's performance. We show that the audit does not necessarily placing a high performance cost on the platform and characterize the conditions under which the platform complies with regulation while remaining  profitable. 
Additionally, we find that, under these same conditions, the platform is not incentivized to create echo chambers if they wish to be profitable.

\section{Background}\label{sec:background}

In this section, 
we discuss the social media regulatory landscape and position our contributions relative to the ongoing discourse, 
focusing in particular on the US. 
We begin by reviewing three factors that moderate lawmakers' ability to regulate social media in the US. 
We then investigate recent suggestions to regulate social media according to flexible, user-driven standards. 
We conclude with an analysis of the benefits and limitations of this approach. 

\subsection{Legal challenges} \label{sec:legal_challenges}

In the US, regulating social media is particularly challenging, 
in part due to three factors: \emph{Section 230 of the Communications Decency Act}, 
\emph{free speech protections}, 
and \emph{trade secret law}.

\textbf{Section 230 of the 1996 Communications Decency Act} protects internet service providers, including social media platforms \citep{brannon2021section}. 
In particular, Section 230(c)(i) states that internet service providers cannot be legally treated as the ``publishers'' or ``speakers'' of any content that they distribute, as long as the provider does not \emph{create} the content. 
Section 230 has effectively given social media platforms immunity when it comes to the content that they host. 
Platforms are (with very few exceptions) allowed to distribute and amplify content without legal repercussions, 
even if the content contains misinformation, hate speech, and the like. 
Moreover, 
Section 230(c)(ii) permits platforms to  \emph{remove} and demote content as they wish, simply requiring that platforms do so in ``good faith'' (an ill-defined concept).

Broadly speaking, there are two regulatory implications of Section 230. 
First, as long as Section 230 stands, 
regulations that use a global standard of acceptable and unacceptable content  (e.g., require the removal of misinformation) are tenuous, 
as platforms cannot be held responsible for the informational content that they distribute.
Second, platforms have largely remained self-regulated in the US due to the strength of Section 230 immunity.
As such, strategies for ``improving'' algorithmic filtering must be mindful of their effect on platforms, as a platform is unlikely to adopt self-regulatory measures that drastically hurt its bottom line.  

In the US,  
platforms are also shielded by the \textbf{First Amendment}, which protects the ability to ``express'' oneself without interference from the government (with a few exceptions, such as in cases of defamation and blackmail).
This right has made it challenging to draw ``bright lines'' between acceptable and unacceptable content.
Consider the following example. 
Suppose that lawmakers pass a law requiring platforms to remove misinformation. 
Unless there is a universally agreed upon definition of ``misinformation'' \emph{and} a way of identifying misinformation with perfect accuracy, 
attempts to comply with this law will likely remove innocent posts because platforms would rather behave conservatively than distribute illegal content \citep{keller2021amplification,1959smith}. 
Because the law unintentionally causes innocent posts to be removed (a phenomenon known as the ``chilling effect''),
it would constitute interference from the government on users' speech. 
Historically, laws that have a tendency to limit  speech (whether intentionally or not) are struck down by the Supreme Court (cf. \emph{Smith v. California} \citep{1959smith}).

On top of all this, 
filtering algorithms are hidden behind \textbf{trade secret law}, 
hindering the government's ability to investigate how the algorithms work and when they are (or are not) to blame. 
In the US, 
any technology that a company attempts to keep confidential and brings the company economic benefit when it remains
confidential qualifies as a trade secret. 
In rare cases, 
lawmakers may grant access to trade secrets (e.g., if a company's actions are shown to hurt consumers). 
However, filtering algorithms are currently hidden from both researchers and lawmakers. 
As a consequence, (i) regulations should not depend on the specific algorithm used to filter content
and (ii) audits should work with only black-box access to a filtering algorithm. 

The approach that we adopt in this work is designed to be compatible with the three legal considerations discussed above. 

\subsection{Defining ``acceptable'' content is difficult}

As discussed in the previous section,
designing content moderation regulations is challenging because it is unclear what behavior lawmakers should deem ``acceptable.''
Creating a strict, global standard of acceptability is often infeasible. 
For example, 
a regulation that prohibits the distribution of misinformation or hate speech generally conflicts with both Section 230 and free speech protections (see the discussion in Section \ref{sec:legal_challenges}).
Under Section 230, a platform cannot be held responsible for as the ``publisher'' or ``speaker'' of content that it distributes, i.e., the platform is not liable for \emph{the information contained in the content it hosts}. 
Moreover, under the First Amendment, bright line regulations are unlikely to survive the US Supreme Court due to their tendency to infringe on the free speech of users and content creators 
\citep{1959smith,keller2021amplification}.

In recognition of this fact, 
there have been proposals to regulate how platforms amplify (and de-amplify) content \citep{cobbe2019regulating,gillespie2022not}.
Although de-amplifying a type of content is not as definitive as prohibiting it, 
case law suggests that this approach is also ill-fated \citep{keller2021amplification}. 
In particular, it runs into the same First Amendment roadblocks described above.
As a result of this ongoing discussion on social media regulations, many have come to the conclusion that regulations based on 
strict, global standards of ``acceptable content'' are unlikely to survive. 

There is, however, another path forward:
regulating with respect to a \emph{flexible, user-driven} standard. 
That is, 
the criteria with which a platform is evaluated should be user-dependent (i.e., vary across users) and increase user agency (i.e., respect what users indicate that they do and do not want). 
\citet{keller2021amplification} discusses this approach in ``Amplification and Its Discontents,'' stating,
\begin{quote}
	``Finally, options based not on speech regulation [...] 
	may offer some paths forward, bypassing many of the constitutional difficulties described above. 
	These models are, at heart, about
\emph{increasing user agency} [emphasis added]. They would not prevent individuals from actively choosing
	lawful but harmful or polarizing content online, any more than current law prevents that same choice in media consumption. But [...] 
	these approaches could alleviate other important problems relating to online content.''
\end{quote}
In this work, 
we build on this perspective and propose a flexible, user-driven baseline for social media. 
Although flexible baselines have been examined
\citep{bandy2020auditing},
our goal is to provide a general framework for regulating and auditing social media platforms.

\subsection{Implications of our user-driven approach}

In this section, we discuss the benefits and limitations of flexible, user-driven regulations.
We focus in particular on the implications of our approach, 
as described in Sections \ref{sec:intro} and \ref{sec:problem_statement}. 

\paragraph{User agency.}
User-driven regulations increase the accountability of social media platforms to users by requiring that platforms filter according to the users' \emph{expressed} interests. 
User-driven baselines are motivated by the idea that users should have some degree of control not only over the content that they create and the data that they provide, 
but \emph{also over the content that they are shown}.
Platforms should not be able to singlehandedly select user content without an indication of user consent. Although consent is poorly defined in the social media context, 
a user-driven baseline encourages content curation that respects user agency and consent. 
 
Unlike regulations that seek to define and regulate ``harmful'' content (a task
that is notoriously difficult both technologically and legally
\citep{keller2021amplification}), 
a user-driven baseline shifts the power to control content to users.
A user-driven approach stands in contrast to regulations that draw \emph{bright lines}, 
i.e., establish a global standard with which platforms must comply, 
which has several drawbacks.
For one, 
bright lines are drawn by policymakers. 
Instead of empowering users, 
bright lines simply transfer the authority to define ``appropriate'' content from the hands of social media giants into the hands of lawmakers, 
bypassing users entirely.
For another, brights lines are rigid, as discussed next. 

\paragraph{Flexible.}
When designing regulations, 
policymakers must balance specificity against generality. 
Regulations that are too specific risk being applicable in limited contexts or becoming obsolete. 
The approach that we adopt in this work---requiring that filtered and baseline feeds have similar downstream effects, as described in Sections \ref{sec:intro} and \ref{sec:problem_statement}---is flexible and adaptable. 
Because the baseline feed is different for each user, 
it does not impose a global standard across all users. 
This approach is also flexible in that it is algorithm-agnostic. 
It does not require the filtering algorithm to be of a specific form. 
On the other hand, regulations that work only for specific algorithms may not be applicable across platforms or as platforms develop new algorithms.

\paragraph{Compatible with existing laws.} 
The user-driven approach that we propose in this work does not conflict with Section 230, 
free speech protections, 
or
trade secret law. 
Under a user-driven regulation,
platforms are not held responsible for the {information} conveyed in the filtered content. 
In this way, they are not treated as the ``publishers'' of the content that they filter, 
which is compatible with Section 230. 
Rather, a user-driven regulation holds platforms accountable to users. 

In addition, 
this user-driven approach protects the freedom of speech.
In general, free speech protections have made it difficult for lawmakers to regulate social media content directly. 
Even if lawmakers agree on a definition of ``harmful'' content, 
criminalizing 
its distribution often results in the \emph{chilling effect}: 
to play it safe, 
platforms over-police content and end up removing innocent posts (e.g., articles are mistakenly labeled misinformation). 
If a regulation leads to this outcome, 
courts will likely strike it down for infringing on the free speech of content creators whose innocent posts are removed. 
A user-driven approach, on the other hand, 
does not define universally ``good'' or ``bad'' content. 
Rather, it places agency into the hands of users, 
allowing them to control what they see.
As articulated by \citet{keller2021amplification},
\begin{quote}
	``This approach avoids many First Amendment problems, because it does not
	involve government preferencing of content. Internet users who lost reach and audience under such a law would have little basis to object, since that loss would stem
	from other users’ choice not to listen to them. Indeed, the Supreme Court’s examples of narrower tailoring in prior cases about communications technologies often
	involved exactly this: letting individuals decide what content they want to see, rather than putting that decision in the hands of an intermediary or other centralized
	authority.''
\end{quote}
Lastly, 
as mentioned in Section \ref{sec:intro}, 
the regulatory framework and audit that we propose in this work require only black-box access to a platform's filtering algorithm. 
As such, it allows platforms to keep their algorithms hidden as trade secrets. 

\paragraph{Limitations.} 
One limitation
of user-driven regulations is that what users want (or indicate that they want) may not lead to desirable outcomes. 
For instance,
a user may enjoy conspiracy theories, 
which would be reflected in their baseline feed. 
Requiring that a filtered feed be sufficiently similar to a baseline feed generally implies that the filtered feed also contain conspiracy theories. 
The approach described in this work (and user-driven regulations, more broadly) would therefore not prevent platforms from showing conspiracy theories to that user. 
As such, 
user-driven regulations are not be-all, end-all solutions, and 
we recommend that they stand alongside other complementary regulations. 
For instance, 
if there is an outcome that lawmakers universally agree is undesirable, 
a user-driven approach alone is unlikely to prevent it. 
Even so, 
\citet{keller2021amplification} supports user-driven regulations, 
stating, 
``If our behavior—our ``revealed preferences,'' in economic parlance—says we want trashy but legal content, should laws prevent platforms from giving it to us?'' 
Although allowing users to select their own content can lead to undesirable outcomes, 
restoring user agency is an important step towards correcting the power imbalance between platforms and users.

One may also wonder whether user-driven content may create and even magnify echo chambers. 
Although echo chambers may arise, 
we show in Sections \ref{sec:audit}-\ref{sec:simulations} that the regulatory framework we propose does not incentivize platforms to create them. 

\section{Problem statement}\label{sec:problem_statement}

In this section, 
we present a regulatory framework. 
Under this framework, 
the feed that a platform filters for a given user must be ``similar'' to a baseline feed that would be produced for the same user.
To make this precise, 
we adapt the notion of decision robustness presented by \citet{cen2021regulating}. 
Intuitively, 
two feeds are 
decision-robust if the informational content contained in both feeds are sufficiently similar such that a user would make similar downstream decisions, 
regardless of which feed they were given. 
Importantly, 
although the auditor tests for decision robustness, 
we require that they do so without needing to access or infer private user information.

Decision robustness has several benefits. 
First, it ensures that the downstream outcomes of two feeds are similar, as desired. 
Second, because it focuses on the downstream outcomes, 
it is possible for the two feeds to contain different content, 
which ultimately proves important to showing that the regulatory framework does not necessarily place a high performance cost on the platform.

\begin{figure*}[t]
	\includegraphics[width=\textwidth]{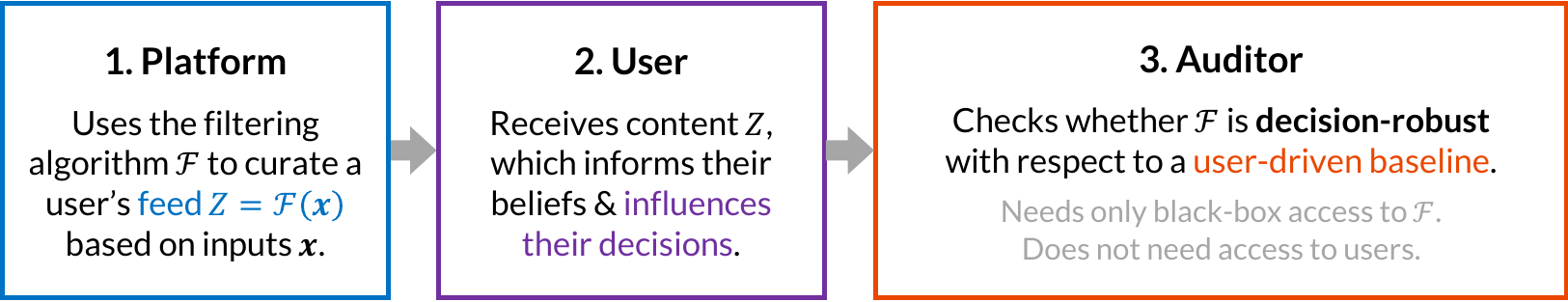}
	\centering
	\caption{Consider a setup with three agents: platform, user, and auditor. 
	The platform employs a filtering algorithm $\cF$. 
	Based in inputs $\bx$ (e.g., the available content and the interaction history of the user), 
	the platform filters the feed $Z = \cF(\bx)$. 
	(For simplicity, we focus on a single feed at a single time step, but our setup can be generalized to multiple time steps). 
	The content $Z$ that the user views may affect the user's beliefs and, as a result, may influence the user's decisions. Note that, if content did not influence users' beliefs and decisions, then we would not be concerned with content moderation to begin with.
	Finally, the auditor monitors the platform by checking whether $\cF$ is decision-robust with respect to a user-driven baseline, as described in Section \ref{sec:DR}. 
	Notably, the auditor should not have more than black-box access to $\cF$, and the auditor should not need to access or infer private user information.}
	\label{fig:setup}
\end{figure*}

\subsection{Setup}

Consider a system with three agents: 
a platform, a user, and an auditor.  
In order to curate feeds for its users, 
the \textbf{platform} employs a filtering algorithm $\cF : \cX \rightarrow \cZ^m$. 
The filtering algorithm takes in inputs $\bx \in \cX$, 
which could encode anything from the history of user behaviors  (e.g., what they have clicked on or watched in the past) to the set of available content.\footnote{The set of available content may change with time. As such, we could write $\cZ$ as $\cZ_t$. For simplicity, we use $\cZ$, which can be easily generalized to the time-varying setting.} 
The filtering algorithm then produces a feed $Z = \cF(\bx)$ for said \textbf{user}, 
where $Z = (\bz_1, \bz_2, \hdots, \bz_m)$ is a feed containing $m$ pieces of content, 
and $\bz_i \in \cZ$ denotes the $i$-th element in the user's feed. 

We restrict our attention to a single feed, 
although it is straightforward to extend our analysis to multiple feeds (e.g., an auditor may wish to examine the feeds across days).

\begin{assumption}\label{asm:iid}
	Let $Z = (\bz_1, \bz_2, \hdots, \bz_m) = \cF(\bx)$ for some $\bx \in X$.  
	Then, $z_{i} \iid p_\bz(\cdot ; \theta(\bx))$  for  $\theta(\bx) \in \Theta \subset \bbR^r$. 
\end{assumption}
\noindent For notational simplicity, 
we will abbreviate $\theta(\bx)$ to $\theta$, when the input $\bx$ is clear from context. 

The \textbf{auditor}'s goal is to check whether the feed that a platform filters for the given user is ``similar'' to the user's baseline feed. 
However, there are two constraints on the auditor:
\begin{enumerate}
	\item The auditor is given only \emph{black-box access} to $\cF$. 
	By ``black-box access'', 
	we mean that the platform can run $\cF$ on a set of $n$ inputs $X = (\bx_1, \bx_2, \hdots, \bx_n)$ and observe the outputs $(Z_1, Z_2, \hdots, Z_n)$, 
	but the auditor does not have additional information about $\cF$.\footnote{The number of times $n$ that the auditor can query $\cF$ is usually limited by the fact that, if the auditor had infinite queries, it would be able to reconstruct $\cF$, 
		which would violate trade secret protections. 
		We do not study this aspect of the audit in this work. 
		We assume that $n$ is given.}
	
	\item The auditor would also like to respect \emph{user privacy}.
	As a result, $\bx_i$ need not correspond to real users. 
	They could, instead, 
	correspond to synthetic users or  denote user features that have undergone a privacy-preserving pre-processing step. 
	
\end{enumerate} 

\subsection{Baseline feed}\label{sec:formal_contract}

Recall the notion of a \textbf{baseline feed} from Section \ref{sec:intro}.
The baseline feed represents the content that a user agrees, or {consents} \citep{cen2021regulating,ghosh2019new,bennett2010regulating}, to viewing. 
There is often a natural choice of baseline feed (e.g., the chronological timeline is often used as a baseline for Twitter \citep{huszar2022algorithmic,bartley2021auditing}). 

As motivated in Sections \ref{sec:intro}-\ref{sec:background},
we build on the perspective that a user's filtered feed should be ``similar'' to their baseline feed at a given time step.
Requiring that a filtered feed is close to a baseline feed aligns with the sentiment that platforms should not have the authority to significantly sway the user beyond what the user consents to. 
In this section, we formalize the notion of a baseline feed and the auditor's objective.  

Formally, 
consider a (hypothetical) user with corresponding inputs $\bx \in \cX$. 
The platform constructs this user's feed from two pools of content:
\begin{enumerate}
	\item We refer to the first pool, $\cZ_B(\bx)$, as the user's \emph{baseline pool}: 
	content to which the user has given consent, 
	such as	updates from the user's friends, articles on pages that the user subscribes to, posts by influencers that the user follows, and so on. 
	
	\item We refer to the second pool, $\cZ_I(\bx)$, as the user's \emph{injected pool}: 
	content that the platform may inject into the user's feed, 
	such as ads, suggested posts, and recommended products.
\end{enumerate}
As such,
the platform constructs the user's filtered feed $Z = \cF(\bx) \subset \cZ_B(\bx) \cup \cZ_I(\bx)$ by selecting some content from $\cZ_B(\bx)$ as well as injecting content from $\cZ_I(\bx)$.

\begin{assumption}
	We assume that the platform provides the auditor black-box access to a baseline filtering algorithm $\cB : \cX \rightarrow \cZ^m$, 
	where for given inputs (i.e., a hypothetical user) $\bx$, 
	$(\bfb_1, \bfb_2, \hdots, \bfb_m) = \cB(\bx)$. 
	As a default baseline, consider a baseline algorithm that draws the content uniformly at random from $\cZ_B(\bx)$, i.e.,
	$\bfb_i \iid \text{UAR}(\cZ_B(\bx))$ for $(\bfb_1, \bfb_2, \hdots, \bfb_m) = \cB(\bx)$.
\end{assumption}
\noindent This assumption is not strong. 
Providing access to a baseline filtering algorithm  adds little extra burden to a platform that is already providing black-box access to their filtering algorithm $\cF$. 
Under this setup, 
the auditor's goal can be expressed as follows: 
\begin{quote}
	\vspace{4pt}
	\emph{Given a set of inputs $X = (\bx_1, \bx_2, \hdots, \bx_n)$, 
	the auditor requires that the filtered feed
	$\cF(\bx_i)$ is ``similar'' to the baseline feed $\cB(\bx_i)$
	for all $i \in [n]$. }
\end{quote}
It remains to determine an appropriate notion of ``similarity.'' 
To this end,
we turn to the concept of \emph{decision robustness}.

\subsection{Decision robustness}\label{sec:DR}

\begin{figure*}[t]
	\includegraphics[width=\textwidth]{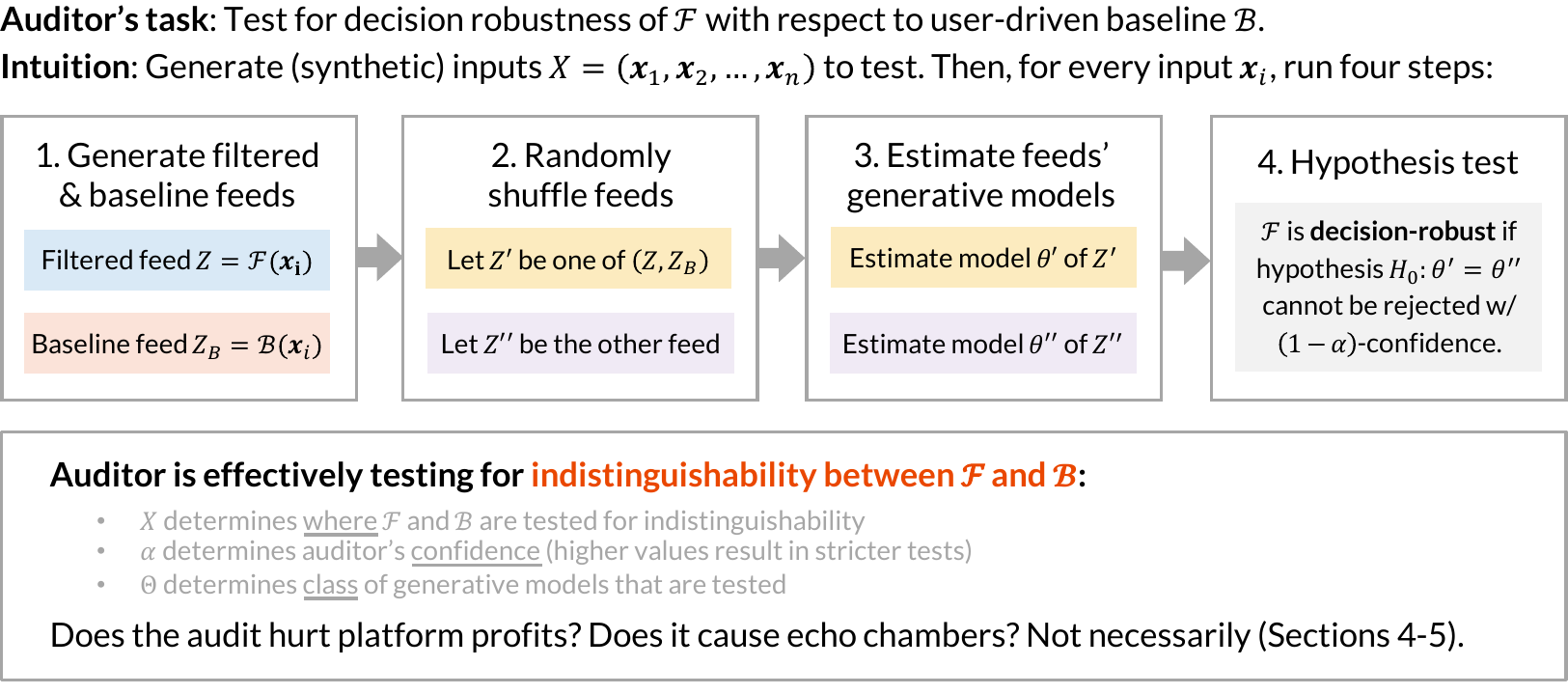}
	\centering
	\caption{The auditor's task is to test for decision-robustness of $\cF$ with respect to a user-driven baseline $\cB$. 
	Intuitively, one can think of this test as follows. 
	Determine inputs $X = (x_1, \hdots, x_n)$ to query (note that these inputs can be generated synthetically and therefore preserve user privacy). 
	For each input, proceed in four steps. 
	First, generate the filtered and baseline feeds $Z = \cF(\bx_i)$ and $Z_B = \cB(\bx_i)$. 
	The baseline $\cB$ could be, e.g., drawing posts uniformly at random from a user's social network (their friends and users that they follow). 
	Second, randomly shuffle the feeds so that $(Z', Z'') = (Z, Z_B)$ with probability $1/2$ and $(Z', Z'') = (Z_B, Z)$ otherwise. 
	Third, assuming that the feeds $Z',Z''$ can be produced by unknown generative models $\theta', \theta'' \in \Theta$, estimate $\theta', \theta''$. 
	Finally, test the hypothesis $H_0 : \theta' = \theta''$. 
	If the auditor cannot reject the null hypothesis $H_0$ with $(1 - \alpha)$-confidence, 
	then $\cF$ is $(\bx_i, \alpha, \Theta)$-decision robust. 
	The auditor is effectively testing for indistinguishability of $\cF$ and $\cB$ in a specific way: \emph{namely, that feeds that they generate do not impart significantly different information, relative to $\Theta$}. 
	Two questions then arise. 
	First, does decision robustness hurt the platform's performance? 
	Second, does indistinguishability result in echo chambers by showing the user more concentrated content? 
	The answer to these questions is no, as examined in Sections \ref{sec:audit}-\ref{sec:simulations}.
	}
	\label{fig:decision_robustness}
\end{figure*}

\emph{Decision robustness} is a concept developed by \citet{cen2021regulating}
that specializes the notion of robustness to the context of algorithmic filtering. 
Intuitively, 
decision robustness is built on the idea that 
the ``similarity'' between two feeds 
should be measured with respect to their downstream impact on users, 
and it uses learning and decision theory to formalize a definition of similarity. 
We slightly modify the definition of decision robustness as follows. 
\begin{assumption}\label{asm:shuffle}
	Once the auditor is given $Z = \cF(\bx)$ and $Z_B = \cB(\bx)$, 
	the feeds are randomly shuffled such that the auditor does not know which feed is the baseline feed and which is the filtered feed. 
	Let the shuffled (unidentified) feeds be denoted by $Z'$ and $Z''$, 
	i.e., $(Z', Z'') = (Z, Z_B)$ with probability $1/2$ and 
	$(Z', Z'') = (Z_B, Z)$, otherwise. 
	By Assumption \ref{asm:iid}, 
	the auditor assumes that 
	$z'_i \iid p_\bz(\cdot ; \theta')$ 
	and
	$z''_i \iid p_\bz(\cdot ; \theta'')$ for all $i \in [m]$ 
	and some $\theta', \theta'' \in \Theta \subset \bbR^r$.  
\end{assumption}
\begin{definition}[Decision robustness for an input]\label{def:DR}
	Suppose Assumptions \ref{asm:iid}-\ref{asm:shuffle} hold.
	For a given $\bx \in \cX$, 
	let  $Z'$ and $Z''$ be as defined in Assumption \ref{asm:shuffle}.
	$\cF$ is $(\bx, \alpha, \Theta)$-decision robust 
	if and only if 
	the uniformly most powerful unbiased (UMPU) hypothesis test with significance $\alpha$ cannot reject the hypothesis $H_0: \theta' = \theta''$.
\end{definition}
\begin{definition}[Decision robustness for a set of inputs]
	Consider a set of inputs $X = (\bx_1, \bx_2, \hdots, \bx_n)$.  
	$\cF$ is $(X, \alpha, \Theta)$-decision robust if and only if 
	$\cF$ is $(\bx, \alpha, \Theta)$-decision robust for all $\bx \in X$.
\end{definition}

\noindent
See the Appendix for a precise definition of the UMPU test. 
Intuitively, 
decision robustness implies that no reasonable hypothesis test cannot confidently reject the hypothesis that both feeds are generated by the same $p_{\bz}(\cdot ; \theta') = p_{\bz}(\cdot ; \theta'')$ for $\theta' , \theta'' \in \Theta$. 
In other words, 
$\cF$ is decision robust if $Z$ and $Z_B$ are \emph{indistinguishable} with respect to $\Theta$. 

\paragraph{Interpretation.}
There is a strong intuition behind the notion of decision robustness that we summarize below. 
Consider a (hypothetical) user who is shown the filtered feed $Z = \cF(\bx)$, 
then faces a set of decision points $Q$. 
The decision points $Q$ captures any possible decision point that the user may face, 
such as where the user decides to eat that evening, 
what shoes to buy, 
and even which presidential candidate to support.
Note that we consider \emph{any} possible $Q$, 
as long as each decision point has a finite number of options. 

Suppose that, in a parallel universe that is otherwise identical, 
the user is shown $Z_B = \cB(\bx)$ instead of $Z$. 
Since both universes are otherwise identical, 
the user faces the same decision points $Q$ after seeing $Z_B$. 
Let the decisions that the user makes in the first universe be denoted by $D$ and the latter by $D_B$.  
As $Q$ is identical in both universes, 
any differences between $D$ and $D_B$ can be attributed to differences $Z$ and $Z_B$ (excepting differences due to randomness).

Suppose we had access to $D$ and $D_B$. 
(Recall that this thought experiment is hypothetical, 
and the audit we propose does \emph{not} need observations of the user's decisions, which may be unethical to observe or infer.) 
Suppose we are \emph{not} told whether $D$ represents the decisions from the filtered or the baseline universe, 
and similarly for $D_B$. 
Then, decision robustness has the following intuition (cf. Section 2 in \citep{cen2021regulating} for details). 
\begin{quote}
	\vspace{4pt}
	\emph{If $\cF$ is $(\bx, \alpha, \Theta)$-decision robust, 
	then (under mild conditions)
	the auditor cannot say with $(1 - \alpha)$-confidence that $D$ is from the filtering universe but $D_B$ is not for \underline{any} $Q$.}
	\vspace{4pt}
\end{quote}
Stated differently, 
decision robustness ensures that the downstream effect of the filtered feed on a user's (hypothetical) decisions---no matter what decision points $Q$ they face---is similar to the downstream effects that the baseline feed would have had.
\begin{example}
	As a toy illustration, 
	consider two coins. Suppose that the first coin has bias $\theta_1$ and the second coin has bias $\theta_2$. Suppose that the platform flips both coins $n = 10$ times,
	the first sequence if observed by user 1, 
	and the second sequence is observed by an \emph{identical} user 2. 
	Suppose that user 1 estimates $\theta_1$ based on the sequence they observe, then uses the estimate $\hat{\theta}_1$ to make decisions $D_1$ in response to $Q$.
	For example, let $Q$ contains only binary decisions, and user 1 choose the first option of each decision point with probability $\hat{\theta}_1$.
	Suppose that user 2 behaves analogously. 
	
	Then, decision robustness holds if $D_1$ and $D_2$ are sufficiently similar such that one cannot confidently say that the two coins have different biases, i.e., that $\theta_1 \neq \theta_2$. 
	It reflects the intuition that the filtered feed should impart similar information as the baseline feed, 
	where similarity is measured with respect to the feeds' effects on downstream decisions. 
\end{example}

Notably, 
decision robustness does \emph{not} require that the filtered feed $\cF(\bx)$ is identical to the baseline feed $\cB(\bx)$. 
Rather, it only requires that they are similar in their downstream effects. 
This flexibility is critical, 
as it allows the platform some freedom in how they filter---the source of their revenue---and it allows platforms to inject personalized content to the benefit of users.

\subsection{Auditor's objective}\label{sec:auditor_objective}

Given the definition of decision robustness, 
checking whether the platform's filtered feeds are similar to their respective baseline feeds can be translated as follows:
\begin{center}
	$\cF$ complies
	$\iff$ 
	$\cF$ is $( X, \alpha, \Theta )$-decision robust
\end{center}
As such,
given a set of inputs $X$, false positive rate $\alpha$, and model family $\Theta$, 
the auditor's goal is to determine whether $\cF$ is $(X, \alpha, \Theta)$-decision robust with
(i) no more than black-box access to $\cF$ 
and 
(ii)  without needing to access or infer private user information.  
We assume that $X$, $\alpha$, and $\Theta$ are given. 
We provide intuition for $X$, $\alpha$, and $\Theta$ in the next section;
how to select them is beyond the scope of this work.

\section{Auditing the algorithm}\label{sec:audit}

In this section, we present an auditing procedure. 
We first introduce notation and definitions, 
then unpack the auditing procedure in detail. 
We conclude with two results. 
The first states that, if $\cF$ passes the audit, the filtering algorithm $\cF$ is asymptotically, approximately decision robust (i.e., that the platform's filtered feeds are close to their respective baseline feeds).
The second shows that the audit does not necessarily place a high performance cost on the platform---that is, the platform can pass the audit with little to no effect on its ability to generate revenue.

\subsection{Notation and definitions}

Let $\chi^2_r$ denote the chi-squared distribution with $r$ degrees of freedom. 
Let $\chi^2_r ( a )$ be defined such that $P( u \leq \chi^2_r ( a ) ) = a$,
 where $u \sim \chi^2_r$. 
Let $I(\theta) \in \mathbb{R}^{r \times r}$ denote the Fisher information matrix at $\theta$ (see the Appendix for a precise definition). 
Let $Y = (\by_1, \by_2, \hdots, \by_q) \in \cY^q$. 
Suppose that $\by_i \iid p_{\by}(\cdot \, ; \phi)$ for all $i \in [q]$, 
where $\phi \in \Phi$ is unknown. 
An \emph{estimator} $\cE : \cY^* \rightarrow \Phi$ produces an estimate $\cE(Y)$ of the parameters $\phi$ that generated $Y$ \citep{lehmann2006theory}.
\begin{definition}[Maximum likelihood estimator] \label{def:MVUE}
	Suppose $\by_i \iid p_{\by}(\cdot \, ; \phi)$ for all $i \in [q]$, as described above. 
	When it exists, 
	the \emph{maximum likelihood estimator}  (MLE) $\cE^+ :  \cY^* \rightarrow \Phi$ 
    satisfies
	\begin{align*}
		\prod_{i=1}^q p_\by( \by_i ; \cE^+(Y) ) \geq 	\prod_{i=1}^q  p_\by( \by_i ; \phi )
	\end{align*} 
	for all $\phi \in \Phi$.
\end{definition}

\begin{algorithm}[t]
	\SetAlgoLined
	\KwIn{Inputs $X \in \cX^n$; 
		black-box access to the filtering algorithm $\cF : \cX \rightarrow \cZ^m$;
		black-box access to the baseline algorithm $\cB : \cX \rightarrow \cZ^m$;
		model family $\Theta \subset \bbR^r$;
		regulation parameter (or false positive rate) $\alpha \in [0,1/n]$.
	}
	\KwOut{$\textsc{pass}$ if $\cF$ is $(X, \alpha, \Theta)$-decision robust; $\textsc{fail}$ if not.}
	
	\vspace{2pt}
	$\tau \leftarrow \frac{2}{m} \chi^2_r ( 1 - \alpha )$\;
	\For {$\bx \in X$} {
		$Z \leftarrow \cF(\bx)$\;
		$Z_B \leftarrow \cB(\bx)$\;
		$(Z', Z'') \hspace{-1pt} \leftarrow \hspace{-1pt} (Z, Z_B) \text{ with probability} \frac{1}{2} \text{ and $(Z_B, Z)$ o.w.}$\;
		$\hat{\theta}' \leftarrow \text{MLE of $\theta' \in \Theta$ given } Z'$\; 
		$\hat{\theta}'' \leftarrow \text{MLE of $\theta'' \in \Theta$ given } Z''$\; 

		\If{$(\hat{\theta}' - \hat{\theta}'')^\top I ( \hat{\theta}' ) (\hat{\theta}' - \hat{\theta}'' )
			\geq \tau$ 
			\textbf{ \emph or }
			$(\hat{\theta}' - \hat{\theta}'')^\top I ( \hat{\theta}'' ) (\hat{\theta}' - \hat{\theta}'' )
			\geq \tau $ \label{lin:test}}{
			Return \textsc{fail}\;
		}
	}
	Return \textsc{pass}\; 
	\caption{Auditing procedure}\label{alg:audit}
\end{algorithm}

\subsection{Auditing procedure}

The auditing procedure is given in Algorithm \ref{alg:audit}. 
This procedure---which is adapted from the algorithm in \citep{cen2021regulating}---repeatedly runs three steps.
For every input $\bx \in X$, 
the auditor runs $\cF$ and $\cB$ to obtain the feeds $Z$ and $Z_B$.
The auditor randomly assigns one of them to $Z'$ and the other to $Z''$. 
The auditor then computes the MLE $\hat{\theta}'$ of parameter $\theta' \in \Theta$ given the samples
$Z' = (\bz'_1, \bz'_2, \hdots, \bz'_m)$.
Similarly, the auditor computes the MLE $\hat{\theta}''$ of parameter $\theta'' \in \Theta$ given the samples
$Z'' = (\bz''_1, \bz''_2, \hdots, \bz''_m)$.
Lastly, the auditor computes two statistics then checks whether they exceed the threshold $\frac{2}{m} \chi^2_r ( 1 - \alpha )$ (Line \ref{lin:test}). 
If either exceeds the threshold for any input $\bx \in X$, 
then the platform fails the audit. 
Otherwise, the platform passes the audit.

Notably, 
the audit requires only black-box access to the filtering algorithm $\cF$ and baseline algorithm $\cB$. 
In addition, 
it does not need to infer or access private user information---it can be run on synthetically generated inputs $X$. 
The selection of $X$ is important. 
$X$ determines the inputs at which $\cF$ is evaluated; one could choose them \emph{a priori} or in an online fashion. 
The auditor must also pick a model family $\Theta$.
Choosing $\Theta$ to be large increases the audit's comprehensiveness while choosing $\Theta$ to be small makes it easier to compute the MLE. 
We recommend choosing $\Theta$ to be a family of multivariate Gaussians or an exponential family. 
For the most part, 
the auditor's main degree of freedom $\alpha \in [0, 1/n]$, 
which can be interpreted as the maximum false positive rate of the hypothesis test.\footnote{Although $\alpha$ can be viewed as a ``rate,'' 
	it is possible for the \emph{platform} to verify that $\cF$ passes the audit for a given $X$ with complete certainty. 
	That is, the platform can test whether $\cF$ complies with the regulation to ensure that they pass the audit. 
	In this sense, $\alpha$ simply quantifies the allowed distance between $\hat{\theta}$ and $\hat{\theta}_B$ or, equivalently, $\cF(\bx)$ and $\cB(\bx)$ for all $\bx \in X$.}
The closer $\alpha$ is to $1/n$, 
the stricter the audit; 
 the closer it is to $0$, 
the looser the audit. 
One of the benefits of the audits is that $\alpha$ is interpretable and one can think of
$n \alpha \in [0, 1]$ as the cumulative false positive rate.

\subsection{Audit's guarantee}

The following result shows that the audit enforces asymptotic, approximate decision robustness. 
As such, 
Algorithm \ref{alg:audit} provides a procedure that ensures the filtered and baseline feeds are close for all $\bx \in X$.

\begin{theorem}
	\label{thm:reg_hyp_test}
	Let Assumptions \ref{asm:iid}-\ref{asm:shuffle} hold.
	Consider an input $\bx \in X$, 
	and let $Z'$ and $Z''$ be as defined in Assumption \ref{asm:shuffle}, 
	i.e., 
	$z'_i \iid p_\bz(\cdot ; \theta')$ 
	and
	$z''_i \iid p_\bz(\cdot ; \theta'')$ for all $i \in [m]$ 
	and unknown $\theta', \theta'' \in \Theta \subset \bbR^r$.  
	Let there be two hypotheses:
	\begin{align*}
		H_0 : \theta' = \theta'' ,
		\qquad \qquad 
		H_1 : \theta' \neq \theta'' .
	\end{align*}
	Let $\theta^* = (\theta'  + \theta'') / 2$. 
	Let $\cP = \{ p_{\bz} ( \cdot \hspace{1pt} ; \theta ) : \theta \in \Theta \}$ denote a regular exponential family that meets the regularity conditions stated in the Appendix. 
	If $\hat{H}$ is defined such that $\hat{H} = H_1 $ if and only if
	\begin{align*}
		(\cE^+(Z')  - \cE^+(Z''))^\top I ( \theta^* ) (\cE^+(Z')  - \cE^+(Z''))  \geq \frac{2}{m} \chi_r^2 ( 1 - \alpha ) , 
	\end{align*}
	then $P( \hat{H} = H_{1} | H = H_{0} ) \leq \alpha$ as $m \rightarrow \infty$. 
	If $r = 1$, 
	then $\hat{H}$ is the UMPU test as $m \rightarrow \infty$. 
\end{theorem}

\noindent 
Recall that $\cE^+$ denotes the MLE (Definition \ref{def:MVUE}),
and observe that the hypothesis test $\hat{H}$ defined in Theorem \ref{thm:reg_hyp_test} closely resembles the test in Line \ref{lin:test} of Algorithm \ref{alg:audit}. 

The main difference is that, in reality, $\theta^*$ in Theorem \ref{thm:reg_hyp_test} is unknown. 
As such, 
Line \ref{lin:test} provides a data-driven version of the test $\hat{H}$ in Theorem \ref{thm:reg_hyp_test}.
Therefore, 
if $\cF$ passes the audit, 
it is asymptotically, approximately decision robust. 
In fact, 
when $r = 1$, 
the audit ensures that $\cF$ is \emph{exactly} decision robust as $m \rightarrow \infty$. 
When $r > 1$, 
the audit does not necessarily execute the UMPU test, 
as required by decision robustness (Definition \ref{def:DR}).
As such, the audit ensures that $\cF$ is approximately decision robust as $m \rightarrow \infty$, 
where it is approximate because there is no guarantee that the audit is the UMPU test (determining the UMPU test for $r > 1$ is known to be a hard problem).

\subsection{Performance cost \& content diversity}

We conclude with a result on the audit's performance cost and its effect on the content a platform filters. 
Suppose that the platform seeks to maximize some reward function $R : \cZ^m \times \cX \rightarrow \bbR$. 
$R$ could, for example, measure revenue, user engagement, content quality, or a mix of these and other factors. 
Let $A(\cZ^m , \bx) \subset \cZ^m$ denote the feasible set: 
the set of feeds that passes the audit for input $\bx$. 
Then, the cost of auditing can be defined as follows. 
\begin{definition}
	The cost of auditing for $\bx$ is $$\max_{W \in \cZ^m} R(W, \bx) - \max_{W \in A(\cZ^m, \bx)} R(W, \bx) .$$ 
\end{definition}
\noindent
The following result shows that the audit does not necessarily impose a high performance cost on the platform. 
In fact, 
there are conditions under which there is no performance cost. 
\begin{proposition}\label{prop:cost}
	If $m < \infty$,
	then there are conditions (stated in Appendix \ref{app:cost_proof}) under which the cost of auditing is $0$ for any $\bx$. 
\end{proposition}
\noindent
We state the precise conditions in Appendix \ref{app:cost_proof}, but
the intuition behind this result is straightforward. 
Recall that the audit requires that the filtered feed has similar downstream effects as the baseline feed, 
as measured by decision robustness. 
There is therefore flexibility in how the platform can filter while still passing the audit. 
The easier it is to distinguish between the filtered and baseline feeds (i.e., to tell them apart with respect to $\Theta$), 
the less flexibility the platform has. 

As we explore in the next section, 
one way to make the two feeds less distinguishable is to add a bit of content diversity to the filtered feed. 
That is, in order to filter while complying with regulation,
platforms are \emph{not incentivized to create echo chambers}. 
Rather, in order to pass the audit, it is in the platform's interest to ensure that every user's feed has sufficient content diversity. As a toy example, even if a user has indicated only interest in conspiracy theories in the past, the platform is incentivized to filter a feed that contains differing theories or information. The conditions stated in Appendix \ref{app:cost_proof} formalize this relationship. 

In order to understand why the audit encourages content diversity, 
consider the following intuition. 
Recall that the platform passes the audit if the auditor cannot confidently determine that $\theta' \neq \theta''$ based on $Z'$ and $Z''$. 
Suppose that there was low content diversity in $Z'$ and $Z''$, 
e.g., all of the news articles in $Z'$ are conservative-leaning while all of the news articles in $Z''$ are progressive-leaning.
Then, an auditor will likely conclude that $\theta' \neq \theta''$.
If, on the other hand, both $Z'$ and $Z''$ have sufficient content diversity, then an auditor cannot confidently reject the hypothesis $\theta' = \theta''$. 
That is, $Z'$ can still be right-leaning while $Z''$ is left-leaning as long they have enough overlap. 
In the audit, \emph{content diversity is quantified by the Fisher information matrices on Line \ref{lin:test}}.

\section{Simulations}\label{sec:simulations}

\begin{figure*}[t]
	\centering
	\begin{subfigure}[b]{0.22\textwidth}
		\centering
		\includegraphics[width=\textwidth]{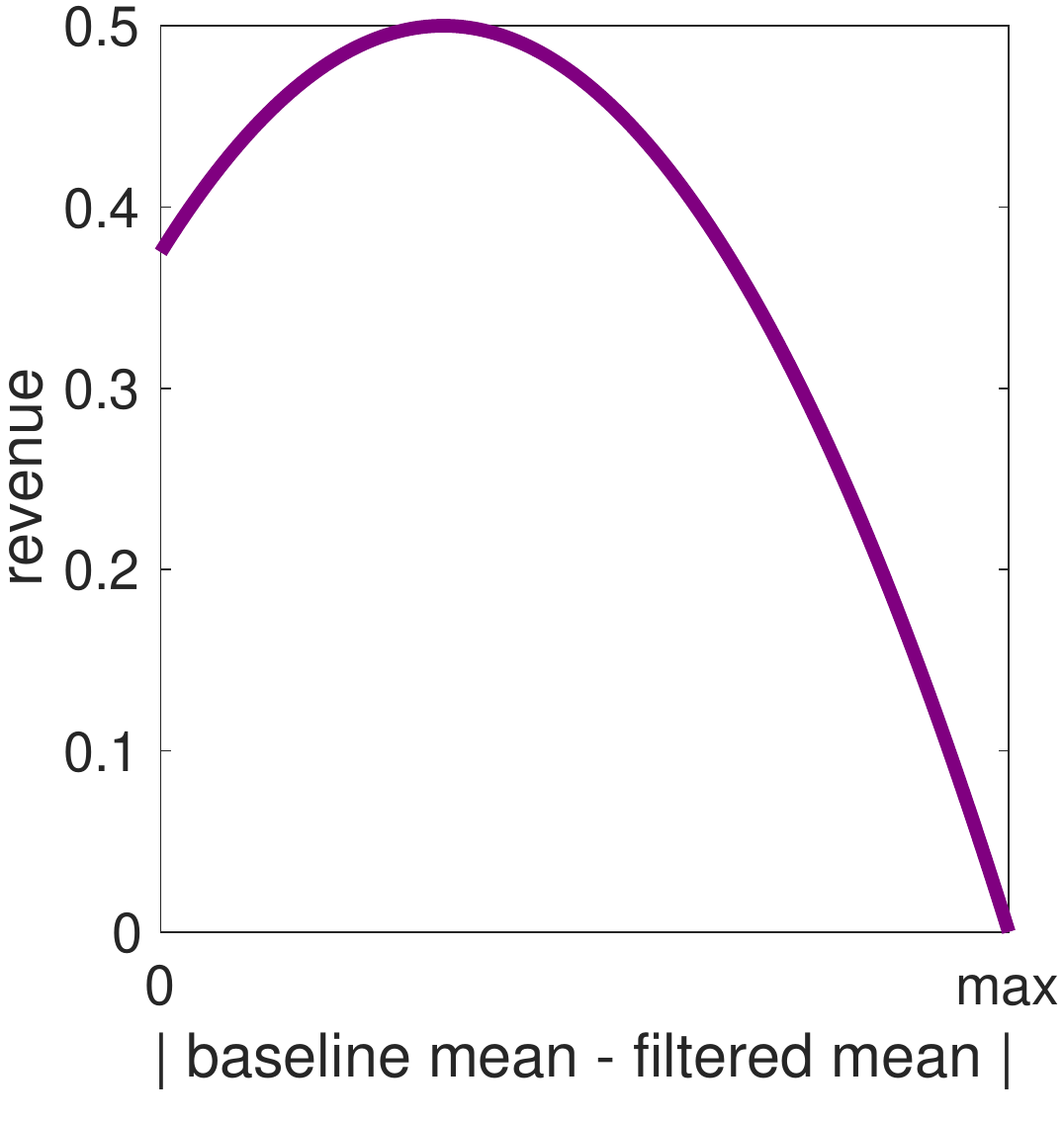}
		\caption{}
		\label{fig:y equals x}
	\end{subfigure}
	\hfill
	\begin{subfigure}[b]{0.284\textwidth}
		\centering
		\includegraphics[width=\textwidth]{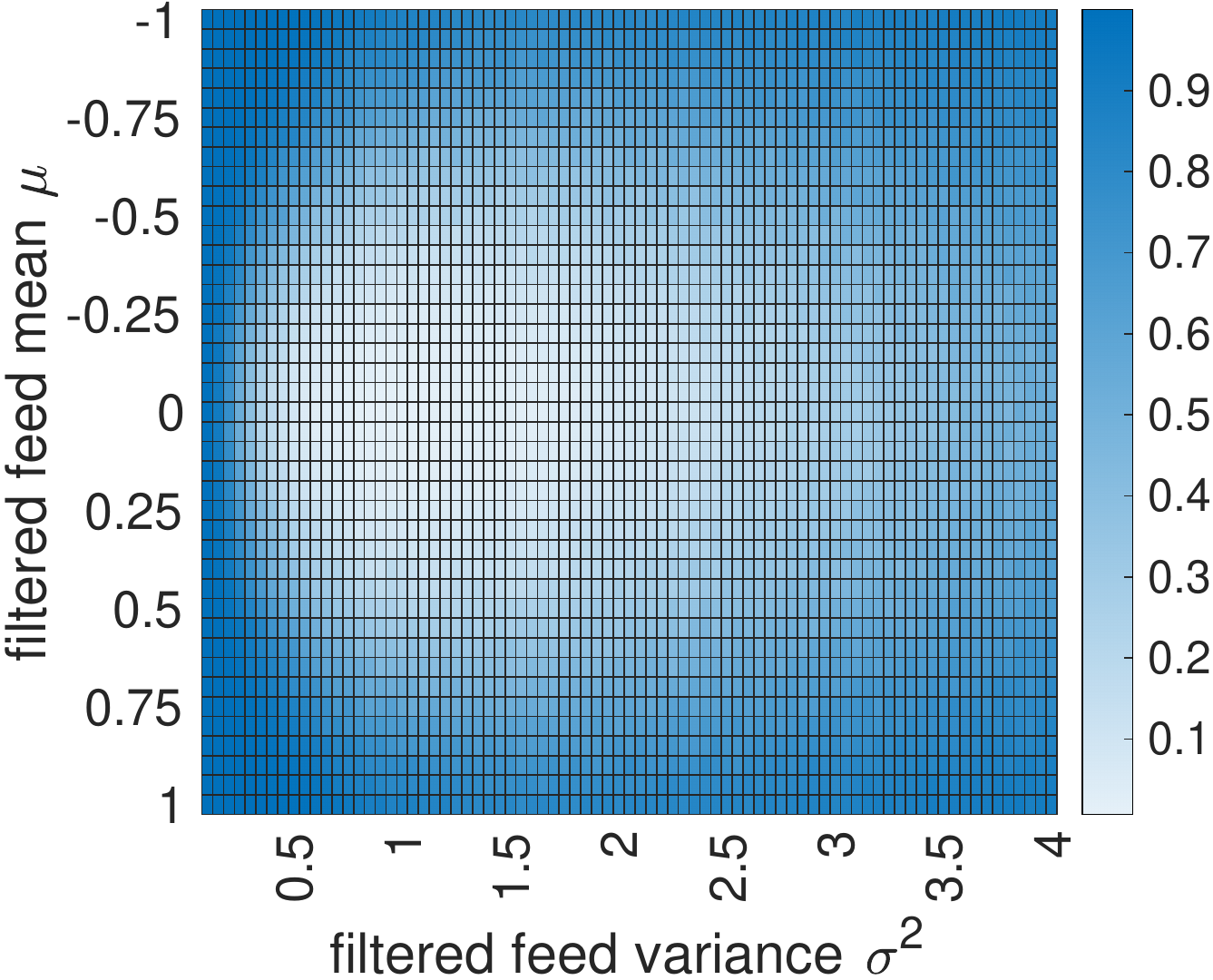}
		\caption{}
		\label{fig:three sin x}
	\end{subfigure}
	\hfill
	\hspace{4pt}
	\begin{subfigure}[b]{0.237\textwidth}
		\centering
		\includegraphics[width=\textwidth]{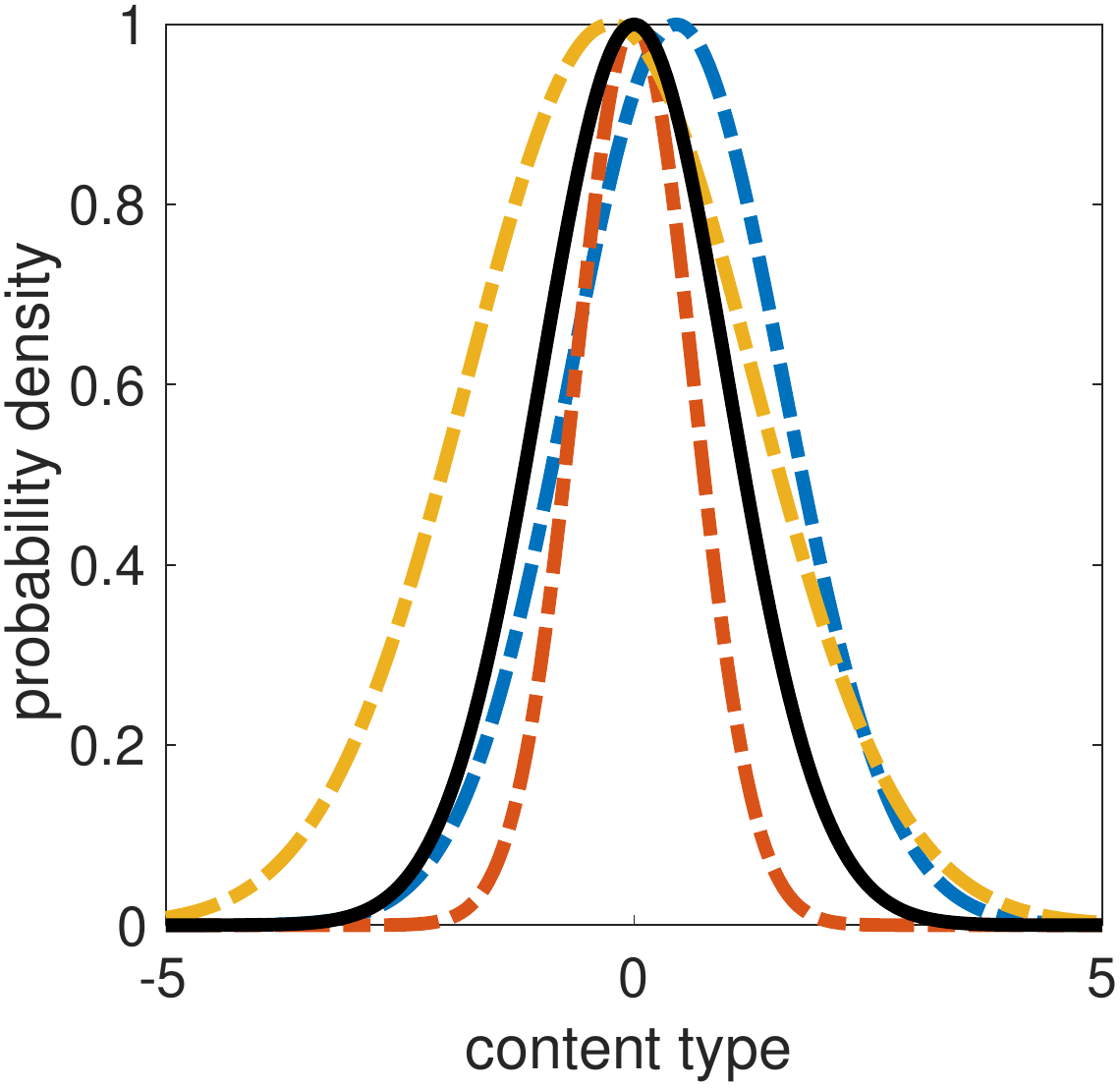}
		\caption{}
		\label{fig:five over x}
	\end{subfigure}
	\hfill
	\begin{subfigure}[b]{0.223\textwidth}
		\centering
		\includegraphics[width=\textwidth]{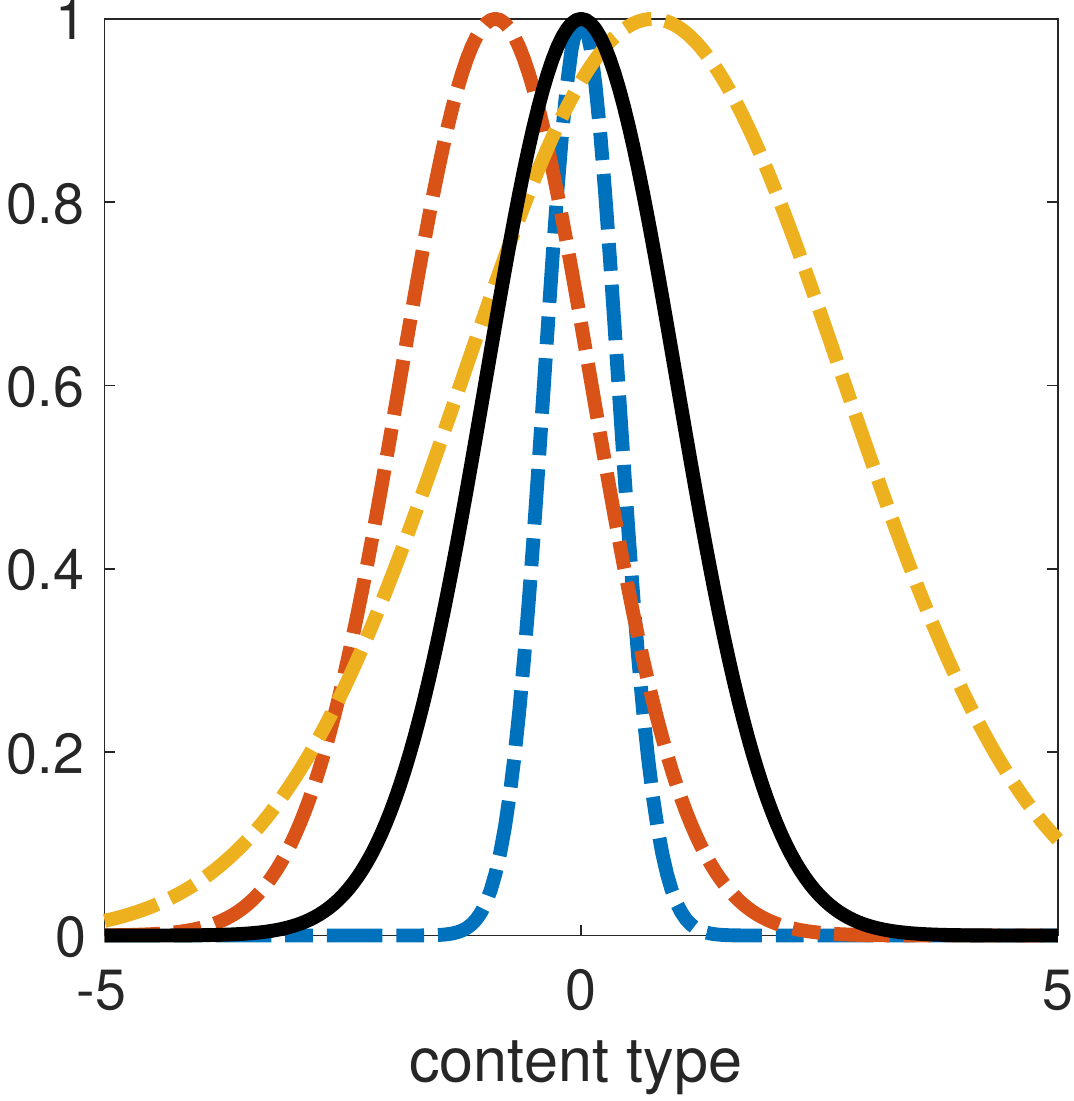}
		\caption{}
		\label{fig:4}
	\end{subfigure}
	\caption{Illustration of the audit and its effect on users' feeds. 
		In (a), we plot an example revenue function that we use in our discussion. 
		This example captures instances where revenue is concave in the distance between the center of the baseline and filtered feeds. 
		Suppose the baseline content follows a standard normal distribution (mean $0$ and variance $1$). 
		In (b), we plot a heatmap, where cell $(\sigma^2, \mu)$ indicates the proportion of filtered feeds drawn from $\cN(\mu, \sigma^2)$ that fail the audit (i.e., are decision robust with respect to the baseline feed $\cN(0, 1)$)
		Lighter cells indicate passing the audit with higher frequency, and vice versa for darker cells.
		In (c) and (d), we illustrate this intuition.
		Both (c) and (d) plot the distribution of the baseline content in solid black. 
		(c) plots three filtered feeds that \emph{pass} the audit in dotted lines (specifically, the distributions used to generate filtered feeds that pass the audit). 
		Similarly, (d) plots three feeds that fail the audit. 
	}
	\label{fig:simulations}
\end{figure*}

In this section, 
we provide simulations that illustrate the effects of auditing.
In particular, 
we examine what types of feeds pass the audit, 
what the platform is incentivized to filter, 
and therefore how the audit changes users' feeds. 
We discuss how the platform is always incentivized to ensure that the filtered feeds contain  diversity (e.g., in terms of viewpoints and across topics) and how this plays nicely with existing work on the importance of content diversity. 

\paragraph{Setup.}
For simplicity, suppose that $\cZ = \bbR$,
i.e., each piece of content is described by a 1-D feature. 
As a toy example, 
suppose that a platform only shows posts estimating today's local temperature.

Now, consider a specific platform and user (i.e., consider a single input $X = \{ \bx \}$). 
We are interested in studying how decision robustness constrains the platform's filtering algorithm $\cF$ and, consequently, affects the filtered feed that is ultimately shown to the user. 
To this end, suppose that the user's baseline content follows a normal distribution $\cN(0, 1)$. 
Recalling that the filtered content must be decision robust with respect to (i.e., ``similar'' to) the baseline content,
we examine what distributions $\cN(\mu, \sigma^2)$ yield feeds that passes the audit. 
In other words, 
suppose the platform composes the user's filtered feed by, first, deciding on parameters $(\mu, \sigma^2)$, 
then drawing content i.i.d. from $\cN(\mu, \sigma^2)$. 
When do such feeds pass the audit? 

We answer this question in our discussion below. 
For the following analysis, let
$\Theta = \bbR \times \bbR_{\geq 0}$, 
$\alpha = 0.01$, 
and
$m = 30$. 

\paragraph{Feeds that pass the audit.}
In Fig. \ref{fig:simulations}(b), 
we give a heatmap, in which each cell's color indicates the proportion of  filtered feeds drawn from $\cN(\mu, \sigma^2)$ that fail the audit over 1000 simulations. 
As expected, 
the rate of failure increases as the filtered content $(\mu, \sigma^2)$ moves away from $(0, 1)$.

Fig. \ref{fig:simulations}(c) plots the baseline distribution (in solid black) and three filtered distributions that generated feeds that passed the audit more than 80 percent of the time. 
Similarly, Fig. \ref{fig:simulations}(d) plots the baseline distribution (in solid black) and three filtered distributions that generated feeds that failed the audit more than 80 percent of the time. 
Although filtered feeds that differ significantly from the baseline feed fail, 
Fig. \ref{fig:simulations}(c)-(d) demonstrate that the platform still has flexibility in how it filters under the audit.

Interestingly, 
however, the heatmap is \emph{not} symmetric.  
Building on the intuition developed by \citep{cen2021regulating}, 
ensuring that the downstream effects of two feeds are (effectively) indistinguishable---in other words, ensuring decision robustness---is easier when the content is less concentrated (i.e., less peaky). 
This observation is reflected in Line \ref{lin:test} of Algorithm \ref{alg:audit}. 
Specifically, 
when the Fisher information is low (i.e., the content is diverse),
the statistic on the left-hand side is small. 
Holding everything else equal, 
low Fisher information therefore reduces the likelihood of failing the audit.

\paragraph{The role of content diversity.}
Consider the revenue function plotted in Fig. \ref{fig:simulations}(a). 
In this simple setup, 
revenue is a strictly concave function of the distance between the means of the filtered and baseline feeds. 
This relationship is built on the following intuition. 
First, the platform earns some baseline amount of revenue by showing the user content that is close to the user's baseline content. 
Second, as the platform shows the user content that is close to but different from their baseline content, the revenue increases---after all, if platforms earned their maximum revenue by filtering baseline content, 
then there would be no need to develop sophisticated filtering algorithms. 
Lastly, at some point, content that is too far from the user's baseline content is no longer relevant to the user, so the revenue that the platform obtains decreases. 

As illustrated by Fig. \ref{fig:simulations}(b), 
ensuring the filtered content is sufficiently diverse can improve the platform's revenue. 
Specifically, consider the column in Fig. \ref{fig:simulations}(b) corresponding to $\sigma^2 = 1.1$. 
As one moves upwards or downwards from $\mu = 0$, 
the proportion of audit failures increases. 
However, for a given $\mu$ (say, $0.5$), 
the platform can increase the pass rate by slightly increasing the filtered feed variance. 
To understand why, 
recall that decreasing the Fisher information improves the platform's chances of passing the audit. 
In the context of one-dimensional Gaussians, 
lowering the Fisher information corresponds to increase variance. 

Intuitively, decreasing the Fisher information---i.e., increasing the filtered feed's variance $\sigma^2$---increases the allowable distance between means (as long as $\sigma^2$ is not too far from 1).
As such, the platform is \emph{incentivized} to include a bit of content diversity in order to maximize its revenue while passing the audit. 
This result shows that the audit plays nicely with efforts to increase content diversity as a remedy for bursting filter bubbles, countering misinformation, and reducing political polarization.

\section{Conclusion}

By moderating the content that users see, 
social media platforms wield a great deal of influence. 
Although lawmakers in the US are eager to regulate social media, 
many efforts to do so are impeded by existing laws and protections. 
In particular, proposals to directly regulate harmful content or regulate platforms' content policies run into problems with Section 230 \citep{brannon2021section} and free speech protections \citep{keller2021amplification}.

In this work, 
we study the effect of algorithmic filtering on users. 
To quantify the effect of algorithmic filtering, 
we use the notion of a baseline feed. 
We then evaluate the similarity between a platform's filtered feeds and the respective baseline feeds using a recently developed concept known as decision robustness. 
Building on recent perspectives to regulate with respect to a user-driven baseline, 
we provide a general regulatory framework. 
We then propose an audit that checks whether a platform's algorithm complies with the regulatory guideline. 
We show that the audit ensures that a platform's filtered feeds are ``close'' to their respective baseline feeds, as desired, 
and that the audit does not necessarily impose a performance cost on the platform. 
We conclude with simulations.

\section*{Acknowledgements}
The authors would like to extend a huge thank you to Aspen Hopkins, Andrew Ilyas, Smitha Milli, and Zachary Schiffer for their feedback, thoughts, and support.

\bibliographystyle{chicago}
\bibliography{ref.bib}

\clearpage

\appendix

\section{Definitions}\label{app:tech_details}

\begin{definition}[Fisher information matrix]
	For a family of distribution $\{ p_{\bz} (\cdot \, ; \, \theta) : \theta \in \Theta \}$, 
	the \emph{Fisher information matrix} $I(\theta) \in \mathbb{R}^{r \times r}$ at $\theta \in \Theta$ is a positive semi-definite matrix, where
	\begin{align*}
		[I(\theta)]_{ij} = \mathbb{E}_{{\sf \mathbf{z}} \sim p_\bz (\cdot  \, ; \,  \theta)} \left[ \frac{\partial}{\partial \theta_i} \log p_\bz ({\sf \bz} ; \theta)  \frac{\partial}{\partial \theta_j} \log p_\bz ({\sf \bz} ; \theta)  \right] .
	\end{align*}
\end{definition}
\noindent 
Recall from Assumption \ref{asm:iid} that $Z$ comprises $m$ samples $\bz_i \in \cZ$ drawn i.i.d. from $p_\bz (\cdot \hspace{1pt} ; \theta)$, for $\theta \in \Theta$. 
\begin{definition}[Asymptotic normality and efficiency]
	An estimator 
	$\cE: \cZ ^m \rightarrow \Theta$
	 is \emph{asymptotically normal and efficient} if:
	\begin{align*}
		\sqrt{m} \left( \cE (Z) - \theta \right) \stackrel{d}{\rightarrow} \mathcal{N}( \mathbf{0}_r , I^{-1}(\theta) ) ,  
	\end{align*}
	as $m \rightarrow \infty$ for all $\theta \in \Theta$ where $I^{-1}(\theta)$ denotes the inverse of the Fisher information matrix at $\theta$.
\end{definition}

\begin{definition}[Uniformly most powerful test]\label{def:ump}
	Suppose that $\hat{G} \in \{G_0, G_1\}$ denotes a binary hypothesis test, 
	where
	\begin{align*}
		G_0 : \phi = \phi' ,
		\qquad 
		G_1 : \phi \neq \phi' ,
	\end{align*} 
	for $\phi, \phi' \in \Phi$.
	Let $G$ denotes the true hypothesis.
	If $\hat{G}$ solves
	\begin{align*}
		\max {\hspace{-1.5pt}}_{\hat{G}'} \hspace{2pt}  P(\hat{G}' = G_1 | G = G_1 )  \quad \text{s.t.}  \quad  P(\hat{G} = G_1 | G = G_0) \leq \alpha , 
	\end{align*}
	for all $\phi, \phi' \in \Phi$ 
	(i.e., maximizes the true positive rate while having a false positive rate---or significance level---no greater than $\alpha$ for all $\phi, \phi' \in \Phi$), 
	then it is the \emph{uniformly most powerful (UMP)} test.
\end{definition}
\begin{definition}[Unbiased test]
	Consider the setup in Definition \ref{def:ump}.
	A hypothesis test $\hat{G}$ is \emph{unbiased}
	if $P(\hat{G} = G_1 | G = G_1) \geq \delta \geq P(\hat{G} = G_1 | G = G_0)$
	for some $\delta \in [0, 1]$.
\end{definition}
\begin{definition}[Uniformly most powerful unbiased test]
	The \emph{uniformly most powerful unbiased (UMPU)} test is the UMP test among all unbiased tests. 
\end{definition}

\section{Regularity conditions}

Let $\cP = \{ p_{\bz} ( \cdot \hspace{1pt} ; \theta ) : \theta \in \Theta \}$. 
The regularity conditions used in Theorem \ref{thm:reg_hyp_test} are given next. 
\begin{enumerate}
	\item $\Theta$ is a compact and open set of $\mathbb{R}^r$.\label{item:reg_cond_1} 
	\item $\bz \iid p_{\sf \bz}(\cdot; \theta)$ for $\theta \in \Theta$ and $\theta_1 \neq \theta_2$ implies $p_{\bz}(\cdot ; \theta_1)$ and $p_{\bz}(\cdot ; \theta_2)$ are distinct.
	\item 
	 The support of $p_\bz(\cdot ; \theta)$ is independent of $\theta \in \Theta$. 
	\item All second-order partial derivatives of $\log p_\bz(\bz ; \theta)$ with respect to $\theta$ exist and are continuous in $\theta$. \label{item:reg_cond_4}
	\item For any $\theta_0 \in \Theta$, there exists a neighborhood of $\theta_0$ and a function $\Pi(\bz)$, where $\mathbb{E}_{{\sf \bz} \sim p_\bz(\cdot \, ; \, \theta_0)} [ \Pi(\bz) ] < \infty$ and 
	\begin{align*}
		\qquad \qquad \left| \frac{\partial^2}{\partial \theta_i \partial \theta_j} \log p_\bz({\sf \bz} ; \theta ) \right| \leq \Pi(\bz) ,
	\end{align*}
	for all $\bz \in \mathcal{Z}$, all $\theta$ in the neighborhood of $\theta_0$, and $i,j \in [r]$.
	\item If $\theta^*$ is the data generating parameter, \label{item:reg_cond_6}
	\begin{enumerate}
		\item $\frac{\partial}{\partial \theta_i} \log p_\bz(\bz ; \theta^*)$ is square integrable for all $i \in [r]$. 
		\item $\mathbb{E}_{{\sf \bz} \sim p_\bz(\cdot ; \theta^*)} \left [ \frac{\partial}{\partial \theta_i} \log p_\bz({\sf \bz} ; \theta^*) \right] = 0$.
		\item Fisher information: 
		\begin{align*}
		\qquad \quad 	[I(\theta^*)]_{ij}  
			&= \mathbb{E}_{{\sf \bz} \sim p_\bz(\cdot ; \theta^*)} \left [ \frac{\partial}{\partial \theta_i} \log p_\bz({\sf \bz} ; \theta^*)  \frac{\partial}{\partial \theta_j} \log p_\bz({\sf \bz} ; \theta^*) \right] 
			\\
			&= -\mathbb{E}_{{\sf \bz} \sim p_\bz(\cdot ; \theta^*)} \left [ \frac{\partial^2 }{\partial \theta_i \theta_j} \log p_\bz({\sf \bz} ; \theta^*) \right] 
		\end{align*}
		\item $I(\theta^*)$  is positive-definite and invertible.
	\end{enumerate}
	\item $\Theta$ is a convex set. 
	\label{item:reg_cond_9}
\end{enumerate} %
As stated in \citep{cen2021regulating},
there are variations on these regularity conditions, 
cf. \citep{bahadur1964fisher,lehmann2006theory,ly2017tutorial}.

\section{Proof of Theorem \ref{thm:reg_hyp_test}}

As stated in the theorem statement, 
Theorem \ref{thm:reg_hyp_test} is adapted from Theorem 1 in \citep{cen2021regulating}.
As such, 
the proof below adapts the proof in \citep{cen2021regulating} as well. 

\begin{proof}
	The definitions and regularity conditions required for Theorem \ref{thm:reg_hyp_test} are given in Appendices \ref{app:tech_details}-\ref{app:tech_details}. 	
	Condition \ref{item:reg_cond_1} (specifically, compactness) and Condition \ref{item:reg_cond_4} ensure that the MLE exists. 
	From Conditions \ref{item:reg_cond_1}-\ref{item:reg_cond_6},
	the MLE is asymptotically normal and efficient \citep{bahadur1964fisher,lehmann2006theory,ly2017tutorial}.
	Under Condition \ref{item:reg_cond_9},
	 $I((\theta' + \theta'')/2)$ exists for $\theta', \theta''  \in \Theta$.
	Under Condition \ref{item:reg_cond_9},
	$I((\theta + \theta')/2)$ exists for $\theta', \theta''  \in \Theta$.
	
	By the asymptotic normality and efficiency of the MLE, 
	\begin{align*}
		\sqrt{m}( \cE^+ (Z')  - \theta' ) &\stackrel{d}{\rightarrow} \mathcal{N}( \mathbf{0}_r , I^{-1} (\theta' )  )
		\\
			\sqrt{m}( \cE^+ (Z'')  - \theta'' ) &\stackrel{d}{\rightarrow} \mathcal{N}( \mathbf{0}_r , I^{-1} (\theta'' )  )
	\end{align*}
	as $m \rightarrow \infty$, where $\bz_i' \iid p(\cdot ; \theta')$ and $Z' = (\bz_1', \hdots, \bz_m')$, 
	and similarly for $Z''$. 
	Therefore, as $m \rightarrow \infty$, 
	\begin{align*}
		\sqrt{m}( \cE^+ (Z')  - \theta' - \cE^+ (Z'') + \theta'' ) \stackrel{d}{\rightarrow} \mathcal{N}( \mathbf{0}_r , I^{-1} (\theta' )  + I^{-1} (\theta'' ) ) 
	\end{align*}
	Under the hypothesis test statement in Theorem \ref{thm:reg_hyp_test},
	$H_0 : \theta' = \theta'' = \theta^*$. 
	Therefore, 
	if $H = H_0$,
	\begin{align}
		\sqrt{m}( \cE^+ (Z')  - \cE^+ (Z'') ) \stackrel{d}{\rightarrow} \mathcal{N}( \mathbf{0}_r , 2 I^{-1} (\theta^*  )  ) \label{eq:asym_norm_diff_F0}
	\end{align}
	as $m \rightarrow \infty$. 
	As a result, 
	the  two-sample, two-sided hypothesis test 
	becomes a two-sample, one-sided test of on the mean of a multivariate Gaussian random variable as $m \rightarrow \infty$. 
	Under \eqref{eq:asym_norm_diff_F0}, 
	\begin{align*}
		(\cE^+(Z')  - \cE^+(Z''))^\top I ( \theta^* ) (\cE^+(Z')  - \cE^+(Z''))  \sim \frac{2}{m} \chi_r^2
	\end{align*}
	Therefore, if $\hat{H}$ satisfies:
	\begin{center}
		$\hat{H} = H_1$
		\\
		$\iff$
		\\
		$(\cE^+(Z')  - \cE^+(Z''))^\top I ( \theta^* ) (\cE^+(Z')  - \cE^+(Z''))  \geq \frac{2}{m} \chi_r^2 ( 1 - \alpha )$ 
	\end{center}
	then $\hat{H}$ has a FPR $\leq \alpha$, as desired.  
	
	$\hat{H}$ as defined above is the UMPU test with significance level $\alpha$ when $r = 1$  (cf. Section 8.3 of \citep{casella2021statistical}).
	As such, 
	$\hat{H}$ would \emph{exactly} test whether $\cF$ is decision robust and the platform upholds the social contract on algorithmic filtering as $m \rightarrow \infty$. 
	When $r > 1$,
	there is no guarantee that $\hat{H}$ is asymptotically the UMPU test (in general, 	determining the UMPU test is known to be difficult for $r > 1$). 
\end{proof}

\section{Proof of Proposition \ref{prop:cost}}\label{app:cost_proof}

\begin{proof}
	Suppose there exists $\Omega \subset [r]$ where $1 < |\Omega| < r$
	and
	$R(\theta_1, \bx) = R(\theta_2, \bx)$ if $\theta_{1,i} = \theta_{2,i}$ for all $i \notin \Omega$.
	Suppose that, for any $\theta \in \Theta$, $\beta > 0$ and $\bv \in \bbR^r$, 
	there exist a vector $\bar{\theta}_{\Omega}$ where $\bar{\theta}_{\Omega,i} = 0$ for all $i \notin \Omega$
	and a constant $\kappa > 0$ such that $\bv^\top I ( \theta + \kappa \bar{\theta}_{\Omega} ) \bv < \beta$ and $\theta + \kappa \bar{\theta}_{\Omega} \in \Theta$. 
	Suppose that $\cZ$ is large enough such that there exist $Z$ and $Z'$ such that $\hat{\theta}$ is the MLE given Z and  $\hat{\theta}{}'$ is the MLE given $Z'$.
	We show that, under these conditions, the cost of auditing is $0$. 
	
	Let $Z_* \in  \arg \max_{W \in \cZ} R(W , \bx)$ be a reward-maximizing feed and $\hat{\theta}_{*}$ be the MLE given $Z_*$. 
	Let $Z_*' \in  \arg \max_{W \in \cZ} R(W , \bx')$ and $\hat{\theta}{}'_{*}$ be the MLE given $Z_*'$.  
	Let 
	$\theta =  \hat{\theta}_{*}$,
	$\beta =  \frac{2}{m} \chi^2_r ( 1 - \epsilon )$,
	and $\bv = \hat{\theta}{}'_{*} - \hat{\theta}_{*}$.
	Finally, let $\hat{\theta} = \hat{\theta}_{*} + \bar{\theta}_{\Omega}$.
	Then, 
	\begin{align*}
		(\hat{\theta}{}'_{*} - \hat{\theta}_{*} )^\top I (  \hat{\theta}_{*} + \bar{\theta}_{\Omega} ) (\hat{\theta}{}'_{*} -  \hat{\theta}_{*} )
		< \frac{2}{m} \chi^2_r ( 1 - \epsilon ) .
	\end{align*}
	Letting $\hat{\theta}{}'  = \hat{\theta}{}'_{*}  + \bar{\theta}_{\Omega}$ and recalling $\hat{\theta} = \hat{\theta}_{*} + \bar{\theta}_{\Omega}$ gives
	\begin{align*}
		(\hat{\theta}' - \hat{\theta} )^\top I (  \hat{\theta} ) (\hat{\theta}{}' -  \hat{\theta} )
		< \frac{2}{m} \chi^2_r ( 1 - \epsilon ) ,
	\end{align*}
	which implies both $Z$ and $Z'$ comply with the regulation.
	Under the stated conditions above, $\cZ$ contains content that is expressive enough such that $Z , Z' \in A(\cZ^m, \bx)$.
	
	It remains to show that the  cost of regulation is $0$. 
	To do so, we show that $Z$, which is in the feasible set, 
	achieves the maximum reward $\max_{W \in \cZ} R(W , \bx) = R(\hat{\theta}_{*} , \bx)$. 
	That the cost of regulation is $0$ follows from the fact that $R (\hat{\theta}_{*}  , \bx) = R(\hat{\theta}_{*}  + \bar{\theta}_{\Omega}  , \bx )$ because $\bar{\theta}_{\Omega,i} = 0$ for $i \notin \Omega$.
\end{proof}

\end{document}